\documentclass[reprint,showpacs,preprintnumbers,amsmath,amssymb, aip,groupedaddress]{revtex4-1}
\usepackage{graphicx,wasysym}
\usepackage{dcolumn}
\usepackage{bm}
\usepackage{tabulary}

\newcommand{\kB}{k_\mathrm{B}}

\begin{document}

\preprint{1}


\title{Illusions of phase coexistence: Comments on ``Metastable liquid-liquid transition ... '' by J. C. Palmer et al., \textit{Nature} \textbf{510}, 385 (2014)\\  }

\author{David Chandler}
\email{chandler@berkeley.edu}
\affiliation{Department of Chemistry, University of California, Berkeley, CA 94720, USA}


\begin{abstract}
The recent paper cited above claims that a molecular simulation of one specific model of supercooled water establishes a stable interface separating two metastable liquid phases, which would imply the existence of metastable two-liquid criticality for that model.  Here, we note that this claim conflicts with fundamental principles and with earlier work published in the \textit{Journal of Chemical Physics}, and we show that the claim is unjustified by the data put forward to support the conclusion.  Other technical problems are also noted.
\end{abstract}

\maketitle

For more than two decades, it has been suggested that anomalous properties of liquid water reflect two distinct liquids and a low-temperature critical point at supercooled conditions.\cite{Poole1992}  Yet Binder has observed\cite{Binder2014} that two-liquid criticality defined in terms of a divergent length scale is impossible at supercooled conditions.  Specifically, growing lengths coincide with growing equilibration times, but the time available to equilibrate can be no longer than the time it takes the metastable liquid to crystallize.  In other words, metastability or instability implies an upper bound to the size of fluctuations that can relax in the liquid.  For water, we will see, this size seems to be no larger than 2 or 3 nm, corresponding to volumes containing fewer than 1000 molecules.
  
This bound is fundamentally different than a cutoff imposed by the practicality of a finite simulation cell.  Transient fluctuations on smaller length scales might seem interpretable in terms of something like a liquid-liquid transition, but the bound implies one can never reach large enough scales to know if that interpretation is correct.  The interpretation certainly seems unnecessary because reasonable molecular models known to not exhibit two-liquid behavior do account for equilibrium anomalies of water\cite{Molinero, Limmer2011, Patey2013}  and nonequilibrium amorphous ices.\cite{Limmer2014} 
 
Nevertheless, two-liquid-like behavior of models of supercooled water remains a curiosity, and Palmer \textit{et al.}~\cite{Palmer2014} claim that one molecular model does exhibit this behavior in a numerical simulation.  The claim is notable in that earlier work\cite{Limmer2013, English2013} would seem to discount the possibility.  It is also notable because Palmer \textit{et al.} suggest that the two-liquid behavior they obtain is close to criticality and can be scaled to large sizes.  Given the bound imposed by metastability, Ref.~\onlinecite{Palmer2014} cannot validate the existence of a critical point in deeply supercooled water.  
Here, several problems are uncovered casting doubt on its conclusions, even those limited to relatively small scales.

First, after reviewing pertinent time scales and length scales intrinsic to supercooled water, we see that it is only the presentation of Ref.~\onlinecite{Palmer2014} that gives the illusion of two-phase metastability.  The data itself does not support the conclusions put forth.  This discussion will be of general interest.  Next, we identify several technical issues pointing to errors in the calculations of Ref.~\onlinecite{Palmer2014}.  That material will likely be of interest to experts only.  The paper ends with a Summary.

\section*{Metastability implies no criticality}

\subsection*{Time scales and length scales}
A metastable state lifetime, $\tau_\mathrm{MS}$, is a property of an irreversible system.  Unlike equilibrium properties, $\tau_\mathrm{MS}$ can depend upon system size and preparation protocols.  For example, upon ordinary cooling to near or below its limit of stability, $T_\mathrm{s} \approx 215$\,K, liquid water will coarsen to ice on time scales as short as $10^{-6}$\,s, while hyper quenching to temperatures well below $T_\mathrm{s}$ at rates comparable to $10^6$\,K/s can produce long-lived glassy states.\cite{Glass}  Further, because no experiment can control all aspects of a nonequilibrium system, $\tau_\mathrm{MS}$ has a distribution of values for any one experimental protocol.  Recent experiments studying ice coarsening in droplets of cold water illustrate this fact.\cite{Nilsson2014} 

Dynamics of deeply supercooled water is heterogeneous, with domain sizes and time scales of growing as temperature, $T$, decreases.  Viewed on small enough length scales and short enough time scales, at the coldest conditions of metastability (i.e., near $T_\mathrm{s}$), these fluctuations are large enough and slow enough to give the illusion of two distinct liquids.\cite{Limmer2013, Yagasaki2014, Limmer2014b}  Not surprisingly, all estimates of the imagined liquid-liquid critical temperature are close to $T_\mathrm{s}$,\cite{Holton2012} but the fluctuations at this temperature are mesoscopic transients characteristic of ice coarsening, not criticality. 

The relevant relaxation time of the liquid, $\tau_\mathrm{R}$, is the time to equilibrate the liquid on length scale $a$, where $a \approx 0.2$ or 0.3 nm is the characteristic microscopic length of the liquid.  This relaxation time is closely related to the metastable lifetime.  Specifically, for $T\lesssim T_\mathrm{s}$, the average and variance of $\tau_\mathrm{R}$ and $\tau_\mathrm{MS}$ grow with decreasing temperature, but the ratio is $\tau_\mathrm{MS}/\tau_\mathrm{R} \lesssim10^3$ throughout this regime.\cite{Limmer2013a}  
Near presumed criticality, the time to equilibrate on length scale $\xi$ would be of order $\tau_\xi=\tau_\mathrm{R} (\xi/a)^z$, where $z \approx 3$.\cite{Hohenberg1977}  But as Binder notes,\cite{Binder2014} $\tau_\xi < \tau_\mathrm{MS}$ because the liquid cannot resist crystallization for times longer than $\tau_\mathrm{MS}$.  Accordingly, $\xi/a< (\tau_\mathrm{MS}/\tau_\mathrm{R})^{1/3}$.  
At coldest metastable conditions (i.e., near where criticality is imagined to appear), it follows that $\xi < 2$ or 3 nm. 

This bound is similar in size to the linear dimension of the simulation boxes used by Palmer \textit{et al.}.\cite{Palmer2014} It is a size that is insufficient to demonstrate the signature of two-phase coexistence, a fact we turn to now.

\begin{figure}
\begin{center}
\includegraphics[width=3.3in]{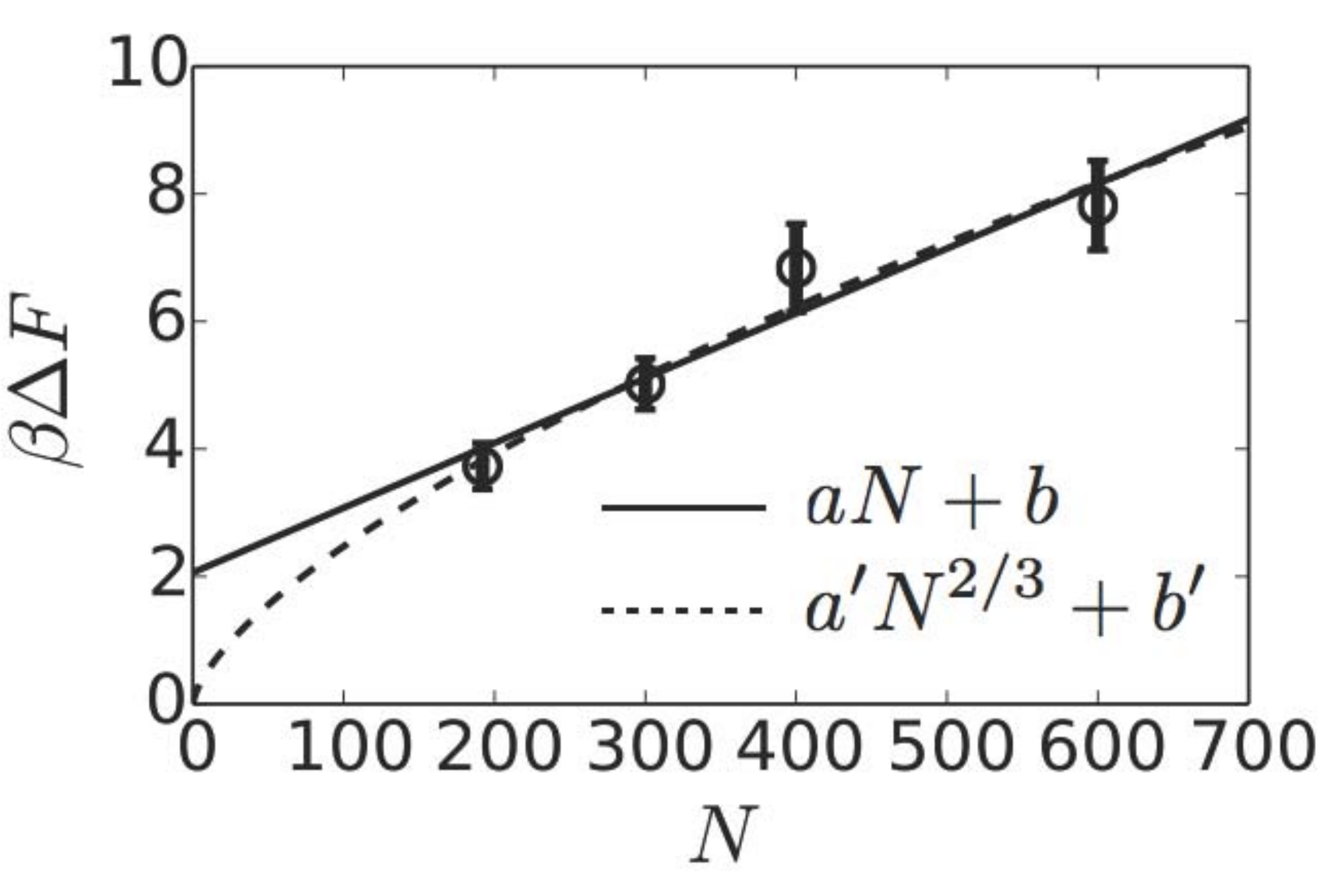}
\caption{ Alleged interfacial free energies as functions of system size $N$.  The black circles and error bars are the four data points from Ref.~\onlinecite{Palmer2014}.  The solid line is the least-square fit of the data to a linear function of $N$.  The dashed line is the least-square fit of the data to a linear function of $N^{2/3}$.  The data cannot distinguish the different functions.}
\label{fig:Fig1}
\end{center}
\end{figure}

\subsection*{No evidence of a stable interface}
Demonstration of two-phase coexistence in a molecular simulation requires evidence of a stable interface separating the two phases.  Thus, for a three-dimensional system grown proportionally in each direction, the interfacial free energy, $\Delta F$, must grow as $N^{2/3}$ where $N$ is the number of molecules in the simulation box.  Palmer \textit{et al.}'s conclusion that their model exhibits a stable interface is based entirely upon showing that four data points as a function of $N^{2/3}$ fall on a straight line to a good approximation.  But the system sizes they consider extend over only a factor of three.  Such a small range is inadequate, as illustrated in Fig.~\ref{fig:Fig1} by least-square fitting the four data points with two different lines.  Both a linear function of $N$ and a linear function of $N^{2/3}$ appear to provide satisfactory fits of the data.   

Thus, it is impossible from the data provided to discern whether or not there is a stable interface.  Other phenomena could be equally or more consistent with the data.  For example, finite transient domains could produce a $\Delta F$ that is initially linear in $N$ and becomes independent of $N$ at large $N$.  

In cases where a first-order phase transition is for certain (e.g., between liquid water and ice), surface scaling can be safely presumed and used to estimate surface tension.  See, for example, Ref.~\onlinecite{Limmer2011}.  But such a presumption is inappropriate for cases where the transition itself is in question.

\section*{Technical issues}

\subsection*{Free energy of the order parameter}
\begin{figure} [b!]
\begin{center}
\includegraphics[width=3.25in]{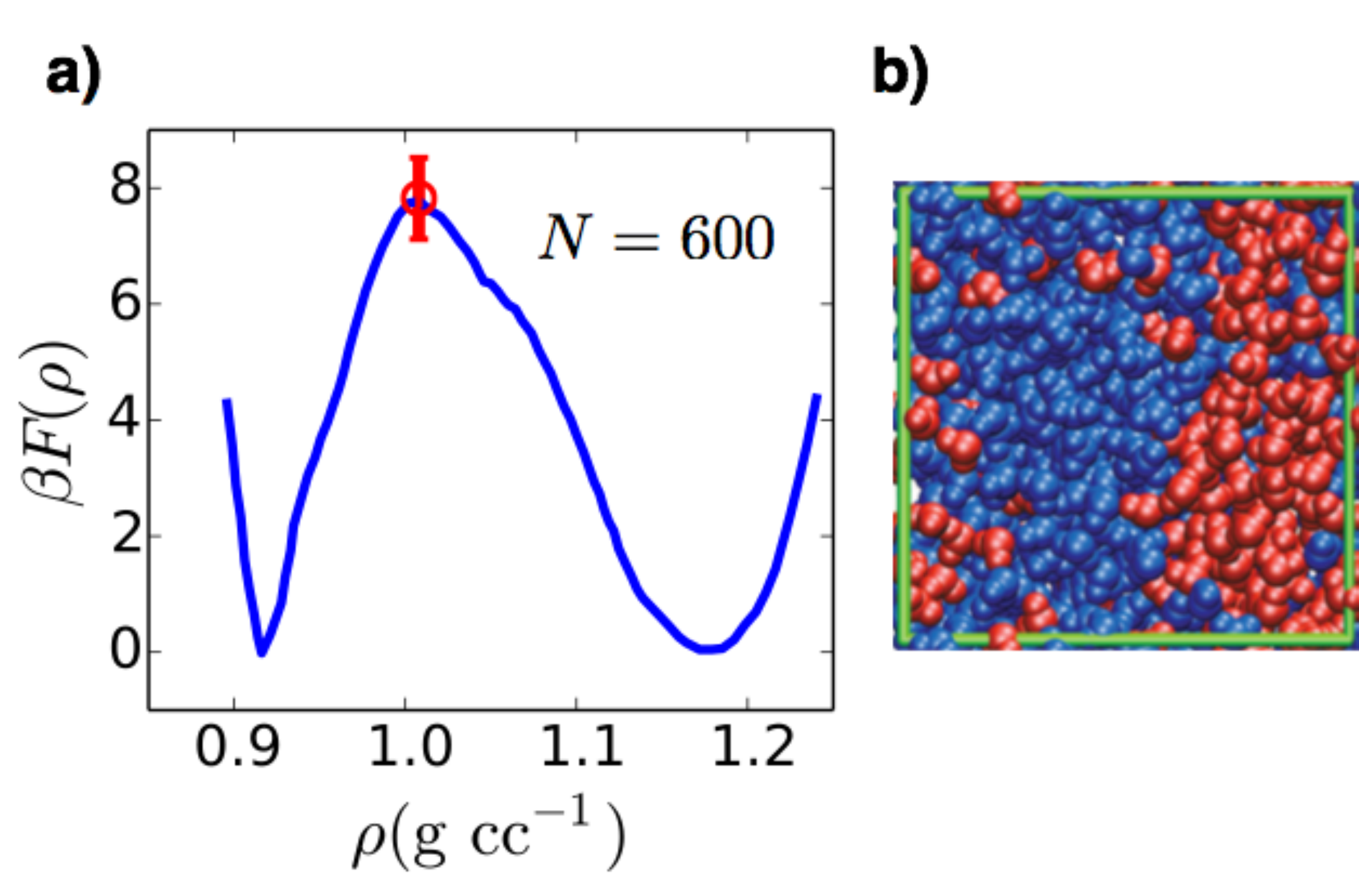}
\caption{(a) Free energy function (in units of $k_\mathrm{B}T = 1/\beta$) calculated and used by Palmer \textit{et al.}\cite{Palmer2014, Private2014} to estimate the surface free energy of the $N=600$ system.  The red circle and error bar show their estimated surface free energy. The anharmonic low-density basin and asymmetric high-curvature barrier suggest poor equilibration and statistical uncertainties much larger than the cited error bar.  (b) A configuration from the simulation of Ref.~\onlinecite{Palmer2014} with $N=600$ at conditions of assumed phase coexistence.  Adapted from Fig. 2 of Ref.~\onlinecite{Palmer2014}.}
\label{fig:Fig2}
\end{center}
\end{figure}

The free energy $\Delta F$ is the reversible work to prepare an interface at conditions of coexistence.  It is essentially the height of the barrier in a bistable free energy as a function of order parameter, which in this case would be the density $\rho$.  This free energy function, $F(\rho)$, is therefore the quantity to be judged when assessing the quality and reproducibility of a calculation of $\Delta F$.   Reference \onlinecite{Palmer2014} does not show this function, but with the information shared with us,\cite{Private2014} Fig.~\ref{fig:Fig2}a graphs Palmer \textit{et al.}'s $F(\rho)$ for their largest system, $N=600$.  Figure~\ref{fig:Fig2}b shows a cross section of the $N=600$ system at the density coinciding with the maximum in $F(\rho)$; it is a snap shot adapted from Fig. 2 of Ref.~\onlinecite{Palmer2014}.  The green box outlines the boundary of the periodically replicated simulation cell.  The blue and red particles are the oxygen atoms of water molecules in high- and low-density regions, respectively.  Reference \onlinecite{Palmer2014} pictures six of the periodic replicas.  Here, Fig.~\ref{fig:Fig2} crops out five of those replicas to avoid a false impression of a more extended interface than observed.   

There are two noteworthy features of the graphed $F(\rho)$:  

1.  The bottom of the left basin is asymmetric, which is contrary to behavior expected of liquid matter.  In particular, reversible fluctuations should yield a parabolic basin up to at least $\kB T$ above the minimum.\cite{IMSMa}  

2.  A bump at the barrier top makes the curvature of the barrier relatively high, which is contrary to behavior expected in the presence of a stable interface.  Specifically, for the configuration pictured in Fig.~\ref{fig:Fig2}b, if the interface were stable, there would be little free energy difference between the pictured configuration and a similar configuration with somewhat thicker red (or blue) regions.  The only curvature at the barrier top should be due to the effects of finite size on fluctuations of the interface, and the barrier top should become flatter as the system size grows.\cite{IMSMb} 

The implication of these features is that the stated error bar largely underestimates the uncertainty and likely error in $\Delta F$.  To the extent curvature at the barrier top is nonzero, there must be deviations from $N^{2/3}$ scaling of the surface energy. The unusual shape of the reported low-density basin suggests that the simulation is trapped in an irreversible state, possibly a glassy configuration.  Indeed, all reasonable models of water will possess such configurations, and it is these configurations that either are or foreshadow the long-lived metastable states of nonequilibrium amorphous ices.\cite{Limmer2014}  

The possibility of errors in free energies due to nonequilibrium effects are generally tested by unweighting neighboring sampling windows and checking for miss alignment and hysteresis.  Either symptom of irreversibility would then be addressed by further sampling.  Reference \onlinecite{Palmer2014} writes that the the ``bootstrap'' method is applied to determine error estimates.  This method is provided by a standard software package for implementing free energy calculations.\cite{MBAR}  It is a reliable error-estimate method provided sampling can be demonstrated to be reversible, which is most easily accomplished if there are no slow variables other than those being controlled in the free energy calculation.  Otherwise, ``bootstrap'' estimates can be misleading. 

The fact that supercooled water can be trapped into nonequilibrium glassy states implies that there are many slow degrees of freedom beyond density, $\rho$, and global order parameter, $Q_6$.  Yet it is only $\rho$ and $Q_6$ that are controlled in the free energy calculations of Ref.~\onlinecite{Palmer2014}, and none of the sampling techniques applied in that work automatically address the concomitant problems of irreversibility.

\subsection*{Time scales}
Free energy functions are reversible work functions.  As such, they should be independent of time scales.  Kinetics enters the picture only to the extent that irreversible dynamics persists in affecting the computed work functions.  There is an infinity of ways by which irreversibility can poison a free energy calculation.  Reference~\onlinecite{Limmer2013} by Limmer and Chandler offers the proposition that apparent bistability often detected by several groups studying supercooled water is the result of slow equilibration of $Q_6$. 

\begin{figure}[t!] 
\begin{center}
\includegraphics[width=3.5in]{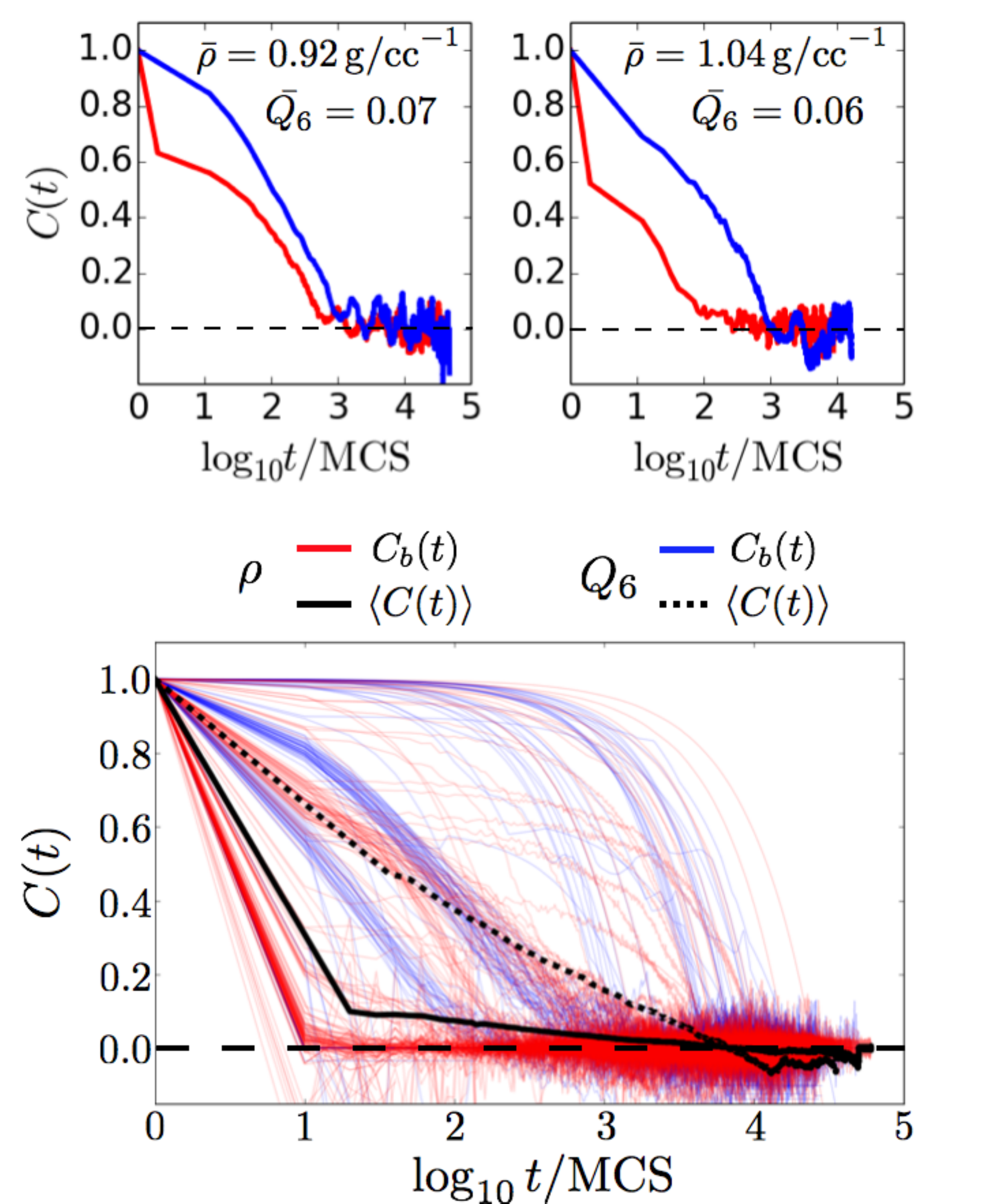}
\caption{ Relaxation functions for a variant of the ST2 model of water at $p=2.2$\,kbar and $T=235$\,K, demonstrating the broad range of time scales and average time-scale separations characteristic of this system.  Red and blue lines are, respectively, the $\rho$ and $Q_6$ autocorrelation functions in different windows sampled during free energy calculations of Ref.~\onlinecite{Limmer2013}.  The functions for two specific windows are shown in the upper two panels, where $\bar{\rho}$ and $\bar{Q}_6$ values specify the averages of the density and global order parameter in the specific window.  Random oscillations about zero at the largest times are the results of autocorrelating over finite times, typically between 50 to 100 times that for the $Q_6$-correlation function to reach 0.1 of its initial value.   Notice two (or more) step relaxation for $\rho$, and that for small $\bar{Q}_6$, the long-time relaxation times of $\rho$ increase and approach those of $Q_6$ as $\bar{\rho}$ decreases.  Averaging all the functions in the lower panel, using the equilibrium weight for each window, yields the black solid and dashed lines in the bottom panel.  The averaged functions are distinct from equilibrium time-correlation functions, but rather illustrate the typical differences between relaxation of $\rho$ and $Q_6$ that must be accounted for to obtain the reversible work surface in $\rho$-$Q_6$ space.}
\label{fig:Fig3}
\end{center}
\end{figure} 

This idea is explored by computing the conditional probability distribution for $\rho$ given a specific value of $Q_6$, $P(\rho | Q_6)$.  Averaging this distribution with the equilibrium distribution for $Q_6$ yields the correct equilibrium free energy, i.e., $\beta F(\rho) = - \ln [\int \mathrm{d} Q_6\, P(Q_6)\, P(\rho | Q_6)]$.   On the other hand, if $Q_6$ is poorly sampled, its distribution will be out of equilibrium, and averaging with that out-of-equilibrium distribution will yield an incorrect free energy function.  Importantly, Limmer and Chandler show,\cite{Limmer2013} if $P(Q_6)$ is fixed at the equilibrium function of the high-density liquid, not letting it adjust to different values of $\rho$, the averaged conditional distribution yields the bistable free energy function reported by Palmer \textit{et al.}\cite{Palmer2014}  In other words, if sampling of $\rho$ proceeds on times scales too short for $Q_6$ to relax, bistable behavior for density will be found.  In contrast, if sampling of $\rho$ and $Q_6$ is sufficient to equilibrate both variables, Limmer and Chandler\cite{Limmer2013} show that the bistable liquid behavior disappears.

Palmer \textit{et al.}\cite{Palmer2014} disagree with this explanation of their earlier results, saying that a time-scale separation between $\rho$ and $Q_6$ does not exist.  This disagreement manifests confusion on at least two levels.  First, the Limmer-Chandler analysis\cite{Limmer2013} is based upon a time-scale separation in the normal high-density region of supercooled water while Ref.~\onlinecite{Palmer2014} considers the low-density amorphous region.  Second, it seems that Ref.~\onlinecite{Palmer2014} incorrectly estimates pertinent relaxation times.

For example, Ref.~\onlinecite{Palmer2014} claims to compute free energy functions by sampling over time scales that are hundreds of times longer than the relaxation time scale for $Q_6$, and it further reports that unconstrained trajectories running for such times show no hint of crystallization.  Such claims contradict findings from 1 $\mu$s molecular dynamics trajectories that exhibit ice coarsening.\cite{Yagasaki2014, Limmer2014b}   The structural relaxation time for a high-density ST2 liquid at the conditions examined is about $10^2$\,ps, and the relaxation time for $Q_6$ is about $10^4$\,ps.\cite{Limmer2013a}  One-hundred times longer would reach $10^6$\,ps, where completed coarsening of the supercooled ST2 model is both observed\cite{Yagasaki2014} and predicted to be observable.\cite{Limmer2013a, Bear}  

It can be difficult to identify absolute physical time scales from a Monte Carlo calculation, such as those carried out for Ref.~\onlinecite{Palmer2014}, but correlation functions can be studied as functions of computation time or Monte Carlo steps.
Figure~\ref{fig:Fig3} shows such relaxation functions for $\rho$ and $Q_6$,
\begin{equation*}
C_b(t) = \frac{\langle \delta \rho(0)\,\delta \rho(t)  \rangle_b}{\langle (\delta \rho)^2  \rangle_b} \quad \mathrm{and} \quad \frac{\langle \delta Q_6(0)\,\delta Q_6(t)  \rangle_b}{\langle (\delta Q_6)^2  \rangle_b}\,,
\end{equation*}
respectively, taken from the calculations of Limmer and Chandler.\cite{Limmer2013}  Here, $\langle \cdots \rangle_b$ denote ensemble average with the biasing potential used to confine configurations to the $b$th window of $\rho$-$Q_6$ space in a free energy calculation.  The fluctuations, $\delta \rho$ and $\delta Q_6$, are deviations from their respective means in the $b$th window.

\begin{figure}[t!] 
\begin{center}
\includegraphics[width=3.3in]{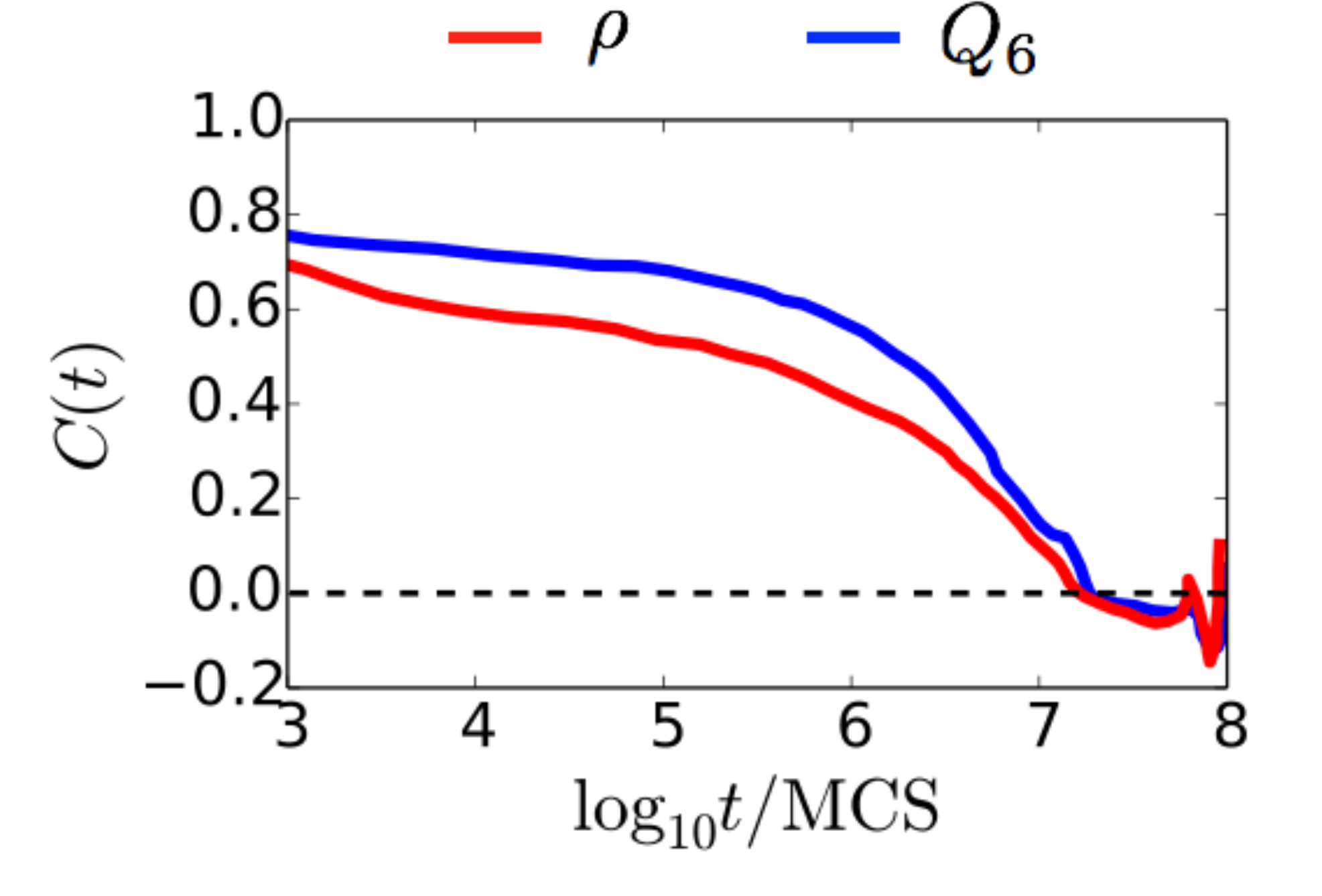}
\caption{ Time-correlation functions for $\rho$ and $Q_6$ computed by Palmer \textit{et al.}\cite{Palmer2014, Private2014} The negative tails indicate a drift in the computations.  Units of time, MCS, are arbitrary, referring to computation steps, and the algorithm for the steps is different from that of Fig.~\ref{fig:Fig3}. }
\label{fig:Fig4}
\end{center}
\end{figure}

Figure \ref{fig:Fig3} shows a broad variety of relaxation behaviors.  The equilibrium weight for a given window, $p_b$, depends upon the thermodynamic conditions under consideration.  The averaged correlation functions, $\langle C(t) \rangle = \sum_b p_b C_b(t)$, are shown in Fig.~\ref{fig:Fig3} for conditions where the normal supercooled liquid is metastable.  The time-scale separation at these conditions is clear.  Moreover, the upper panels of Fig.~\ref{fig:Fig3} illustrates how decreasing density towards that of ice while keeping global order amorphous, the time-scale separation diminishes at the longer times but remains significant at the shorter times.

Reference~\onlinecite{Palmer2014} reports different time dependence for time-correlation functions of $\rho$ and $Q_6$.  The graph provided for these functions in Ref.~\onlinecite{Palmer2014} crops out negative values of $C(t)$, but with information provided to us,\cite{Private2014} a more complete graph is shown here in Fig.~\ref{fig:Fig4}.  These correlation functions were computed by averaging over trajectories initiated in the low density amorphous system -- the part of configuration space where glassy states exist.  It is expected that both $\rho$ and $Q_6$ will relax slowly in this regime.  As such, the correlation functions presented in Ref.~\onlinecite{Palmer2014} do not contradict or discredit the results and arguments put forward in Ref.~\onlinecite{Limmer2013} because Ref.~\onlinecite{Limmer2013} focuses on fluctuations from the normal liquid.

Two features of Palmer \textit{et al.}'s $C(t)$ are noteworthy:  

1.  At the shortest times for which they provide data, $C(t)$ for the density shows a hint of two-step relaxation.  (More data at shorter times could clarify the extent of this feature.)  Such relaxation is common at glass-forming conditions.\cite{BinderKob}  The suggested early-time relaxation would reflect the relaxation processes that dominate at the higher densities, and it would be consistent with the behavior exhibited in the top left panel of Fig.~\ref{fig:Fig3}.   

2.  The negative tails at large sampling time $t$ seems oscillates about a non-zero negative value.  (More data at longer times could clarify the extent of this feature.) A tail oscillating about a non-zero value would imply a systematic drift in the simulations.  This apparent nonstationarity seems to be direct evidence of poor equilibration.  

\subsection*{Freezing}
In prior work focusing on how fluctuations associated with coarsening of ice can be confused with two-liquid behavior,\cite{Limmer2011,Limmer2013} free energy surfaces as functions of $\rho$ and $Q_6$ for various models have been computed.  Three such models are variants of the ST2 model, which are termed the ST2a, ST2b and ST2c models.  These variants differ only in the way long-ranged interactions are treated, and qualitative behaviors of the three variants are similar.\cite{Limmer2013, Digress}  In their new work, Palmer \textit{et al.}\cite{Palmer2014} study freezing of the ST2b version and attempt to compare with results of Ref.~\onlinecite{Limmer2013}.  The attempted comparison leads Palmer \textit{et al.} to claim that the Limmer-Chandler results of Ref.~\onlinecite{Limmer2013} exhibit large unexplained errors.   This claim turns out to be baseless, as discussed now.

Properties of coexistence -- the temperature-pressure locations, the surface tension, and so forth -- are model dependent, and computing these properties from simulation requires significant care in establishing reversibility, coexistence and system-size dependence.  The Limmer-Chandler treatments of the ST2 models in Refs.~\onlinecite{Limmer2011} and \onlinecite{Limmer2013} are less ambitious, with the purposes of establishing the non-existence of a second liquid phase and establishing the presence of a crystal ice basin.  Therefore, much more statistics and smaller error estimates were obtained for amorphous regions than for crystalline regions.  No attempt was made to locate phase coexistence properties for the ST2 model.  Indeed, the phase diagram for freezing the ST2 model is yet to be determined by anyone.  

Palmer \textit{et al.}\cite{Palmer2014} report that coexistence conditions for the ST2 model are known, but that is not true.  
Weber and Stillinger\cite{WeberStillinger} estimated a temperature at which a small spherical cluster will melt, and from that estimate, they suggest, in effect, that the low-pressure melting temperature for an ST2 model is about 300\,K.   By another means, Ref.~\onlinecite{Palmer2014} estimates another melting point to be $273\pm3$\,K at $p=2.6$\,kbar.  Until now, that seems to be the extent of what is known about  melting points of ST2 models.

\begin{figure*}[t!] 
\begin{center}
\includegraphics[width=6.3 in]{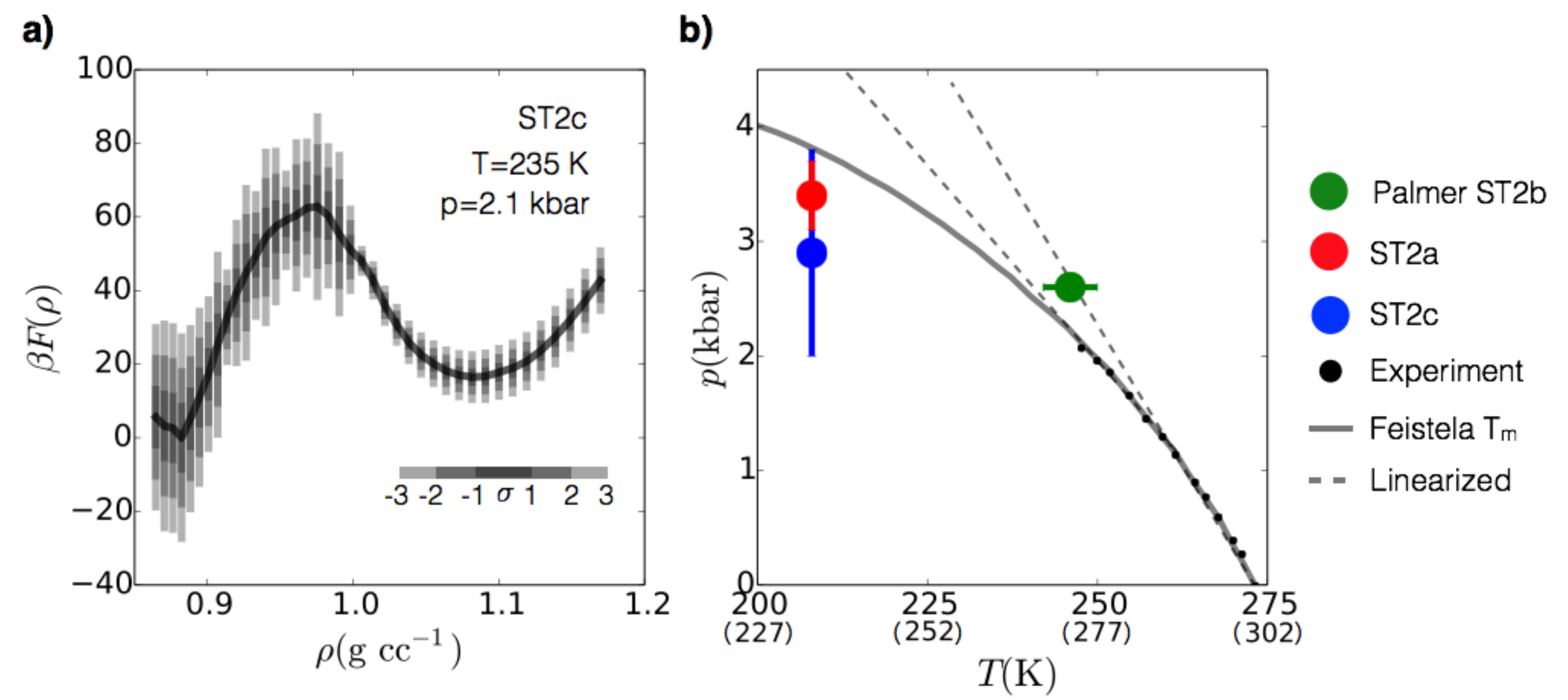}
\caption{Free energy function of order parameter and coexistence line between liquid and ice I. (a) The free energy for the ST2c model as a function of crystal-order parameter $Q_6$, according to the data collected to produce the free energy surface shown in the upper-right panel of Fig. 13 in Ref.~\onlinecite{Limmer2013}; error estimates are for 1, 2 and 3 standard deviations, as indicated. (b)  The water-ice I coexistence line modeled from experimental data by Feistel and Wagner,\cite{Feistel2006} with shifted temperature scale applying to ST2 simulations noted in parentheses.  Linear extrapolations from two experimental points are shown with dashed lines.  The large points locate three estimates of coexistence points extracted from simulation data of three variants of ST2 models at or near pressure $p=2.6$\,kbar or temperature $T=235$\,K.  The red and blue points are deduced from the simulations described in Ref.~\onlinecite{Limmer2013}.  Error estimates for the ST2a variant are from those in Ref.~\onlinecite{Limmer2013}; error estimates for the ST2c variant follow from those shown here in Panel (a).  The green simulation point and error estimate for the ST2b model are from Palmer \textit{et al}.\cite{Palmer2014}  }
\label{fig:Fig5}
\end{center}
\end{figure*}

To augment that limited knowledge, we can glean what information can be obtained from the data presented in Ref.~\onlinecite{Limmer2013}.  With the ST2a model, the statistics is sufficient to locate the coexistence for $T=235$\,K at $p =3.4 \pm 0.3$\,kbar.  This coexistence point is shown here in Fig.~\ref{fig:Fig5}b.  (The relevant free energy function is shown in Fig. 6a of Ref.~\onlinecite{Limmer2013}.)  With the ST2c model, the statistics is less accurate.  See Fig.~\ref{fig:Fig5}a. By re-weighting this free energy function for this variant of the ST2 model, coexistence is found for $T=235$\,K at $p = 2.9 \pm 0.9$\,kbar.  This coexistence point is also shown in Fig.~\ref{fig:Fig5}b.  For the ST2b model, however, the data assembled in Ref.~\onlinecite{Limmer2013} is insufficient to locate a phase coexistence point.  Among other things, the full extents of the ST2b liquid and crystal basins were neither sampled nor shown in Ref.~\onlinecite{Limmer2013}.  

Remarkably, Ref.~\onlinecite{Palmer2014} reports predictions of freezing from the ST2b calculations in Ref.~\onlinecite{Limmer2013}, and it uses these predictions to discredit Ref.~\onlinecite{Limmer2013}.  In actuality, the predictions from the ST2 calculations in Ref.~\onlinecite{Limmer2013} are in reasonable accord with what can be deduced from experiment, assuming Weber and Stillinger's estimate of the low-pressure melting temperature is correct.  In particular, because it is presumed to be 300\,K, the temperature scale for the ST2 model can be imagined to be shifted by about 27\,K from experiment.  With that shift, the experimentally determined phase coexistence line between liquid water and ice I provides an estimate of that for the ST2 model.  The high-pressure portion of that line, $p>2$\,kbar, is an extrapolation into a regime where ice I is metastable with respect to other ice phases.  The equation of state of Feistela and Wagner is used to make that extrapolation, assuming that equation of state remains reliable beyond the regime for which it was derived.  By comparing the extrapolated experimental line with the ST2 estimates in Fig.~\ref{fig:Fig5}b, it appears that that both Ref.~\onlinecite{Palmer2014} and Ref.~\onlinecite{Limmer2013} do equally well (or poorly) in locating coexistence points for the ST2 model of water.

Reference \onlinecite{Palmer2014} also claims that Ref.~\onlinecite{Limmer2013} gives an erroneous value for the chemical potential difference between liquid and crystal at $T=230$\,K and $p=2.2$\,kbar, saying that the results of Ref.~\onlinecite{Limmer2013} predict a value of 66\,J/mol, whereas the correct value is an order of magnitude larger.   
In actual fact, the best estimate from the numerical data of Ref.~\onlinecite{Limmer2013} is about 400\,J/mol.   This estimate uses the ST2a variant and the assumption that all variants will give about the same chemical potential differences.

\subsection*{Summary}
This paper examines much of what Palmer \textit{et al.} have presented in their new publication.\cite{Palmer2014}   Lack of evidence for $N^{2/3}$ scaling in surface free energy is demonstrated.  Signatures of poor equilibration are identified, the most likely explanation being that Palmer \textit{et al.}'s low-density amorphous basin reflects not a low-density liquid, but rather one or more of the low-density states that can contribute to the non-equilibrium glassy states of water.  Finally, criticisms of Ref.~\onlinecite{Limmer2013} are shown to be erroneous.

The main result in contention -- whether a small enough simulation cell of ST2 water exhibits bistability at supercooled conditions -- is not an issue of possible errors in force-field evaluation or algorithm implementation.  Such sources of confusion have been checked against years ago, as has been noted in Ref.~\onlinecite{Limmer2013}.  Rather, disagreement about two-liquid-like bistability is based upon the issue of equilibration or reversibility.  Indeed, Ref.~\onlinecite{Limmer2013} shows that this bistability is reproduced by constraining the $Q_6$-distribution to the standard liquid state distribution, and that this bistability disappears as the $Q_6$-distribution is allowed to relax.  This general result, independent of whatever free energy sampling method is employed, argues that Palmer \textit{et al.}, and others with similar results, need to demonstrate control of equilibration.

Three points require attention:

1.  Unlike that shown in Fig.~\ref{fig:Fig4}, $Q_6$-correlation functions should relax in a fashion consistent with stationary distributions of states. Behaviors like those shown in the upper panels of Fig.~\ref{fig:Fig3} are illustrative of what should be expected.  While relaxation in unconstrained supercooled ST2 water will necessarily exhibit nonstationarity (coarsening near $T_\mathrm{s}$  takes place on time scales that are only $10^2$ longer than an average $Q_6$-relaxation times), constrained ensembles used in free energy calculations should be stationary if sampled for long enough simulation times.

2.  With relaxation of $Q_6$ established in each case, reversibility of free energy calculations should be examined by passing back and forth between neighboring $\rho$-$Q_6$ windows with a variety of different paths.  Sampling within each window should extend for times at least of order $10^2$ larger than those for the $Q_6$-autocorrelation function to relax.  Hysteresis and path dependence will necessarily appear when moving between large and small $Q_6$.   The size of these irreversible effects require consideration in error estimates.   Asymmetry of the low-density basin and sharpness of the barrier top in Fig.~\ref{fig:Fig2} manifest irreversible effects thus far not accounted for by Palmer \textit{et al.}.  These features necessarily imply uncertainties larger by factors of three or more from those reported.  Further, when establishing bistability, the location of coexistence introduces additional errors, which are also not accounted for in Ref.~\onlinecite{Palmer2014}. 

3.  Only if two-liquid-like bistability persists when satisfying Points 1 and 2, would the result meaningfully challenge Ref.~\onlinecite{Limmer2013}.  At that point, one would want to know the limit of length scales for this newly discovered heterogeneity.  For real water, we estimate that any remnant of reversible two-liquid-like behavior will disappear on length scales larger than 2 or 3 nm.  Is the same true for the ST2 model?  Further, what is the physical basis for why the ST2 model at small scales could behave in a fashion that is qualitatively different than other reasonable models of liquid water, and what is the physical basis for why the computations of Ref.~\onlinecite{Limmer2013} miss the effect?  On their disagreement with Ref.~\onlinecite{Limmer2013}, Palmer \textit{et al.}\cite{Palmer2014} offer speculations, but no physical picture with accompanying quantitative results, and the speculations prove to be wrong.

For sure, the length-scale cutoff of metastability implies that the issues in this debate have little to do with large-enough scale simulations or with experimental observations.  Moreover, it is not the same issue as whether multiple amorphous basins exist.  All reasonable models of water exhibit glass-forming states, some of high density, others of low density.  These are  transient states at reversible conditions.  They can become long-lived irreversible states, but only at conditions driven far from equilibrium.  Their systematic exploration therefore requires techniques of nonequilibrium statistical mechanics.  See, for instance, Ref.~\onlinecite{Limmer2014}.     
  
Corresponding-states analysis\cite{Limmer2013a} indicates that ST2 water exhibits precursors to glassy behavior at temperatures significantly higher than those of other models and experiment.  For example, in real water this correlated dynamics begins when temperature is lowered below 277\,K, but in the ST2 model it begins at 305\,K.  It is this higher corresponding-state temperature that seems responsible for the appearance of irreversible artifacts in poorly equilibrated simulations of the ST2 model.\cite{Limmer2013, Limmer2013a}  This is not to say that straightforward but unequilibrated simulations of the ST2 model correctly describe amorphous ices.  Without employing appropriate methods to attend to the enormously longer timescales of glass and glass transitions, erroneous behaviors will be found, and the ST2 model will be mistreated in that way as well.

 \begin{acknowledgments} 
This work was initiated after the publication of Ref.~\onlinecite{Palmer2014}, when I first learned of the paper. Pablo Debenedetti provided simulation data from Ref.~\onlinecite{Palmer2014}, which helped elucidate the contradictions with Ref.~\onlinecite{Limmer2013}.  David Limmer provided generous advice on what I have written here.  My research on this topic has been supported by  the Director, Office of Science, Office of Basic Energy Sciences, Materials Sciences and Engineering Division and Chemical Sciences, Geosciences, and Biosciences Division under the U.S. Department of Energy under Contract No. DE-AC02-05CH11231.
\end{acknowledgments}


\begin{thebibliography}{99}

\bibitem{Poole1992}
P. H. Poole, F. Sciortino, U. Essmann and H. E. Stanley, \textit{Nature} \textbf{360}, 324 (1992).

\bibitem{Binder2014}
K. Binder, \textit{Proc. Natl. Acad. Sci. USA} \textbf{111}, 9374 (2014).

\bibitem{Molinero}
V. Holten, D. T. Limmer, V. Molinero and M. A.  Anisimov,  \textit{J. Chem. Phys.} \textbf{138},174501 (2013).

\bibitem{Limmer2011}
D. T. Limmer and D. Chandler,  \textit{J. Chem. Phys.} \textbf{135}, 134503 (2011).

\bibitem{Patey2013}
S. D. Overduin and G. N. Patey, \textit{J. Chem. Phys.} \textbf{138}, 184502 (2013).

\bibitem{Limmer2014}
D. T. Limmer and D. Chandler, \textit{Proc. Natl. Acad. Sci. USA} \textbf{111}, 9413 (2014).

\bibitem{Palmer2014}  
J. C. Palmer,  F. Martelli, Y. Liu, R. Car, A. Z. Panagiotopoulos and P. G. Debenedetti, \textit{Nature} \textbf{510}, 385 (2014).

\bibitem{Limmer2013}
D.T. Limmer and D. Chandler, \textit{J. Chem. Phys.} \textbf{138}, 214504 (2013).

\bibitem{English2013}
N. J. English, P. G. Kusalik and J. S. Tse, \textit{J. Chem. Phys.} \textbf{139}, 084508 (2013).

\bibitem{Glass}
P. Bruggeller and E. Mayer, \textit{Nature} \textbf{288}, 569 (1980).

\bibitem{Nilsson2014}
J. A. Sellberg, C. Huang, T. A. McQueen, N. D. Loh, H. Laksmono, et al., \textit{Nature} \textbf{510}, 381 (2014).

\bibitem{Yagasaki2014}
T. Yagasaki, M. Matsumoto and H. Tanaka, \textit{Phys. Rev. E} \textbf{89}, 020301 (2014).  This work presents vivid illustrations from a reasonably large-scale molecular dynamics trajectory of the dynamics of ice coarsening.  Its interpretation is discussed in Ref.~\onlinecite{Limmer2014b}.

\bibitem{Limmer2014b}
D. T. Limmer and D. Chandler, arXiv:1406.2651. 

\bibitem{Holton2012}
V. Holten, C. E. Bertrant, M. A. Anisimov and J. V. Sengers, \textit{J. Chem. Phys.} \textbf{136}, 094507 (2012).

\bibitem{Limmer2013a}
D. T. Limmer and D. Chandler, \textit{Faraday Discuss.} \textbf{167}, 485 (2013).

\bibitem{Hohenberg1977}
P. C. Hohenberg and B. I. Halperin, \textit{Rev. Mod. Phys.} \textbf{49}, 435 (1977).

\bibitem{Private2014}
P. G. Debenedetti, private communication (2014).

\bibitem{IMSMa}
D. Chandler, \textit{Introduction to Modern Statistical Mechanics} (Oxford University Press, New York, 1987).  Section 5.3.

\bibitem{IMSMb}
\textit{ibid}.  Pages 174-175.

\bibitem{MBAR}
M. R. Shirts and J. D. Chodera, \textit{J. Chem. Phys.} \textbf{129}, 124105 (2008). https://simtk.org/home/pymbar.

\bibitem{Bear}
Bear in mind that Ref.~\onlinecite{Yagasaki2014} treats a simulation system $10^2$ times larger than considered by Palmer \textit{et al.},\cite{Palmer2014} and the coarsening time to crystalize similar fractions of material at deeply supercooled conditions is larger for the larger system.  But this system size dependence in the deeply supercooled case is weak,\cite{Avrami1939} growing as $N^{1/4}$, so 1\,$\mu$s is not much longer than the time it should take the smaller system to crystalize.  

\bibitem{Avrami1939}
M. Avrami, \textit{J. Chem. Phys.} \textbf{7}, 1103 (1939).

\bibitem{BinderKob}
K. Binder and W. Kob, \textit{Glassy Materials and Disordered Solids} (World Scientific, Singapore, 2005). Chapter 5.

\bibitem{Digress}
Palmer \textit{et al.} have said that the behaviors of the three variants are notably different, and they are unable to converge their calculations on the ST2a variant, but we do not digress on this issue because it is discussed enough elsewhere.\cite{Limmer2013} 

\bibitem{WeberStillinger}
T. A. Weber and F. H. Stillinger, \textit{J. Chem. Phys.} \textbf{80}, 438 (1984).

\bibitem{Feistel2006}
R. Feistel and W. Wagner, \textit{J. Phys. Chem. Ref. Data} \textbf{35}, 1021 (2006)

\end{thebibliography}
\end{document}